\documentstyle[epsf,twocolumn,fleqn]{article}

\begin{document}
\renewcommand{\textfraction}{0.1}
\renewcommand{\topfraction}{0.8}
\rule[-8mm]{0mm}{8mm}
\begin{minipage}[t]{16cm}
\begin{center}
{\LARGE \bf Spin excitations in ferromagnetic manganites\\[4mm]}
J.~Loos$^{\rm a}$ and H.~Fehske$^{\rm b}$\\[3mm]
$^{\rm a}${Institute of Physics, Czech Academy of Sciences, 
CZ-16253 Prague, Czech Republic}\\
$^{\rm b}${Physikalisches Institut, Universt\"at Bayreuth, 
D--95440 Bayreuth, Germany}\\[4.5mm]
\end{center}
{\bf Abstract}\\[0.2cm]
\hspace*{0.5cm}
An effective one--band Hamiltonian for colossal--magnetoresistance (CMR) 
manganites is constructed and the spin excitations are determined. 
Fitting the experimental data by the derived spin--wave dispersion gives 
an  $\rm e_g$--electron hopping amplitude of about $0.2$ eV 
in agreement with LDA band calculations.\\[0.2cm]


\end{minipage}\\[4.5mm]
\normalsize
The theoretical description of the striking magnetic and transport phenomena 
in CMR manganites is far from being satisfactory. There exist  
inconsistencies even in the calculations based on the widely used 
purely electronic double exchange (DE) and ferromagnetic (FM) 
Kondo lattice (KL) models. Indeed, Millis~{\it et al.}~\cite{MLS95} estimated the
Curie temperature $T_c$ from the spin--dependent hopping amplitude of the
DE model and obtained a much higher value than the observed 
one; they ascribed this disagreement to the neglect of the electron--phonon 
coupling. On the other hand, M\"uller-Hartmann and Dagotto~\cite{MD96} 
reexamined the large Hund's rule coupling ($J_H$) limit of the 
KL model and found 
%complex fluctuating electron--bond--hopping amplitude
a nontrivial total bond--spin dependent sign in the effective hopping  
-- arguing that this discrepancy to the DE model  
may be the source of the overestimation of $T_c$.

The purpose of this contribution is to analyse the spin excitation 
spectrum of FM manganites ($\rm [La,A]MnO_3$) on the basis of an 
effective one--band model. Adapting the approach~\cite{LF97} 
to the  $\rm Mn^{3+}$--$\rm Mn^{4+}$ system, the matrix elements 
related to the hopping of an itinerant $\rm e_g$--electron, can be 
determined in the space of spin functions with the  $\rm Mn^{3+}$ spin 
functions restricted to the $S=2$ sub--space 
($J_H\to \infty$)~\cite{comment}. 
Then, focusing on the metallic
FM phase and applying the spin--wave approximation,
the relevant part \mbox{of the effective hopping Hamiltonian is}
\begin{equation}
{\cal H}_{eff} = -t \sum_{\langle i,j \rangle}
\left( C^{\dagger}_{j \frac{1}{2}}  C_{i \frac{1}{2}}^{}
+  C^{\dagger}_{j -\frac{1}{2}}  C_{i -\frac{1}{2}}^{}\right)\,, 
\end{equation}
\begin{eqnarray}
C_{i  \frac{1}{2}}^{}&=& \;|i \mbox{\small $\frac{3}{2}$}\rangle \langle i 2|
+ \mbox{\small $ \frac{\sqrt{3}}{2}$}   
|i \mbox{\small $\frac{1}{2}$}\rangle \langle i 1| \,, \\ 
C_{i -\frac{1}{2}}^{}\!\!\!&=&\!\!  \mbox{\small $ \frac{1}{2}$} 
|i \mbox{\small $\frac{3}{2}$}  \rangle \langle i 1|
+ \mbox{\small $ \frac{\sqrt{2}}{2}$}   
|i \mbox{\small $\frac{1}{2}$}\rangle \langle i 0| \,.
\end{eqnarray}
\begin{figure}[t]
\unitlength1mm
\begin{picture}(70,65)
\end{picture}
\end{figure}
Here the indices $i,\,j$ indicate the positions of $\rm Mn^{4+}$
and $\rm Mn^{3+}$ ions with spin projections $[ \mbox{\small $ \frac{1}{2}$}, 
\mbox{\small $ \frac{3}{2}$}]$  and $[0,1,2]$, respectively.
Based on~(1), the effective transport Hamiltonian in second quantized form
may be derived analogically to~\cite{LF97}. However, 
owing to the strong electronic correlations, the equivalent 
``hole representation'' introduced by Kubo and Ohata~\cite{KO72}
seems to be more suited for the lightly doped manganites. In this picture,
the moving hole with spin $s=1/2$ is strongly antiferromagnetically coupled
to the $\rm Mn^{3+}$ spin background. Consequently, we use  
the Schwinger--boson representation for the $S=2$ spin functions,
and express the basis vectors  of the $S^{\prime}=3/2$ spin space 
%(via Clebsch--Gordon coefficients) 
in terms of fermionic hole operators.
 
As a result we get an effective Hamiltonian for holes interacting with magnons, 
where the ``free'' hole part becomes  
\begin{equation}
{\cal H}^{(0)}_t = \sum_{\bf k} \varepsilon^{(0)}_{{\bf k} \downarrow}
h^{\dagger}_{{\bf k} \downarrow} h^{}_{{\bf k} \downarrow}
\end{equation}
with $ \varepsilon^{(0)}_{{\bf k} \downarrow}=-2 t_h 
(\cos k_x + \cos k_y + \cos k_z)$  and $t_h= [2S/(2S+1)]\, t$.
The interaction terms lead to the spin--dependent hole self--energies 
\begin{eqnarray}
\Sigma_\downarrow ({\bf k},\omega)\!\!\!&=&\!\!\!\frac{1}{4N}\sum_{\bf q} \left[
\left(\varepsilon^{(0)}_{{\bf k}-{\bf q} \downarrow} 
-2 \varepsilon^{(0)}_{{\bf k} \downarrow}\right) n_{\bf q}
\right.\nonumber\\
&&\!\!\!\!+\left. \left(\varepsilon^{(0)}_{{\bf k} \downarrow} \right)^2
\frac{ n_{\bf q} +1 - n_{{\bf k}-{\bf q}\uparrow}}{
\omega-\omega_{\bf q}-\varepsilon_{{\bf k}-{\bf q} \uparrow} +\mu}\right]\\
\Sigma_\uparrow ({\bf k},\omega)\!\!\!&=&\!\!\!\frac{1}{4N}\sum_{\bf q} \left[
\varepsilon^{(0)}_{{\bf k}+{\bf q} \downarrow} n_{\bf q} 
\right.\nonumber\\
&&\!\!\!\!+\left. \left(\varepsilon^{(0)}_{{\bf k}+{\bf q} 
\downarrow} \right)^2
\frac{ n_{\bf q} + n_{{\bf k}+{\bf q}\downarrow}}{
\omega+\omega_{\bf q}-\varepsilon_{{\bf k}+{\bf q} \downarrow} +\mu}\right]
\end{eqnarray}
with $n_{\bf q}\!=\!\left[\exp (\beta \omega_{\bf q}) \!-1\! \right]^{-1}$, 
$n_{{\bf k}\sigma}\!=\!\langle h^{\dagger}_{{\bf k} \sigma} 
h^{}_{{\bf k} \sigma}\rangle $, and 
\begin{equation}
\varepsilon_{{\bf k} \sigma}^{}-\varepsilon_{{\bf k} \sigma}^{(0)}=
\Re\mbox{e} \left[\Sigma_\sigma ({\bf k},\varepsilon_{{\bf k} \sigma}^{}-\mu)
\right]\,.
\end{equation}
The spectrum of elementary spin excitations is determined by 
\begin{eqnarray}
\omega_{\bf q}\!\!&=&\!\!\frac{1}{4N}\sum_{\bf k} \left[
\left(\varepsilon^{(0)}_{{\bf k}-{\bf q} \downarrow} 
-2 \varepsilon^{(0)}_{{\bf k} \downarrow}\right) n_{{\bf k}\downarrow}
+\varepsilon^{(0)}_{{\bf k}+{\bf q} \downarrow} n_{{\bf k}\uparrow} 
\right. \nonumber\\
&&\!\!\!+\left. \left(\varepsilon^{(0)}_{{\bf k} \downarrow} \right)^2
\frac{ n_{{\bf k}\downarrow} - n_{{\bf k}-{\bf q}\uparrow}}{
\varepsilon_{{\bf k} \downarrow}-\varepsilon_{{\bf k}-{\bf q} \uparrow}
-\omega_{\bf q}}\right]\,. 
\end{eqnarray} 
The full solution of the coupled integral Eqs.~(5)--(8) lies outside 
the scope of the present treatment; we estimate the 
spin--wave dispersion at $T=0$ assuming $n_{{\bf k}\uparrow}\simeq 0$,  
$\varepsilon_{{\bf k}\uparrow}\simeq 0$ and 
$\varepsilon_{{\bf k} \downarrow}\simeq \varepsilon^{(0)}_{{\bf k} \downarrow}$.
The latter approximations, justified for small 
hole concentrations $x\ll 1$, are expected to 
give reasonable estimates also for higher doping 
level, provided that the minority spin--up subband 
%$(\sigma=\uparrow)$ 
remains unimportant with respect to the majority 
spin--down subband 
%$(\sigma=\downarrow)$ 
owing to their different spectral weights. 

As a direct test of the theory developed so far, in Fig.~1 
the calculated $\omega_{\bf q}$ has been compared
with recent neutron scattering results for the spin--wave dispersion 
in FM $\rm La_{0.7}Pb_{0.3}MnO_3$ at 10~K. Obviously, the 
dispersion relation~(8) is entirely sufficient to account for the 
measured behaviour throughout the whole Brillouin zone. For $x=0.3$ 
the best least squares fit to the experimental data
fixes the only free parameter of the theory: $t=0.188$~eV.  
Note that this value is in excellent agreement with the result obtained  
from LDA band structure calculations~\cite{SPV96}. Also the consistency of the 
measured magnon bandwidth and $T_c$ was stressed in~\cite{Peea96}.

According to the above estimate of $t$, no polaron band--narrowing is apparent 
at low temperatures. On the other hand, the observed anomalously large oxygen
isotope effect on $T_c$ shows the relevance of the lattice dynamics near
$T_c$, i.e., in the vicinity of the metal--insulator transition~\cite{ZHKM97}. 
To comprise the possible role of polarons, the effective Hamiltonian 
may be easily generalized taking into account the Holstein--type 
interaction of holes with the lattice degrees of freedom. 
However, in order to discuss  the effects near $T_c$ one has
to go beyond the spin--wave approximation. 
\begin{figure}[t]
\centerline{\mbox{\epsfxsize 8.5cm\epsffile{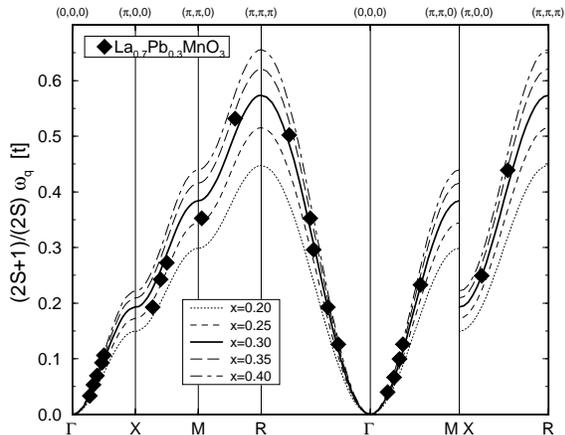}}}\vspace*{-0.5cm}
\caption{Spin--wave dispersion $\omega_{\bf q}$ along all major symmetry
directions of the Brillouin zone compared with the experimental data 
taken from Ref.~\protect\cite{Peea96}, whereby a small constant 
anisotropy gap $\Delta\simeq 2$~meV has been subtracted.}
\end{figure}
 
We are grateful to  A. Wei{\ss}e for numerical support. 
The research was granted by the Czech Grant, 
No. 202/96/0864, and by the Deutsche Forschungsgemeinschaft 
through SFB 279. 
{\small 
\bibliography{ref}

\begin{thebibliography}{10}
\bibitem{comment}
We proved that these matrix elements agree with those
following from Eqs.~(6), (8) given in~\cite{MD96}.
\end{thebibliography}
\bibliographystyle{phys}
}
\end{document}